\input{aipcheck}
\documentclass[
    ,final            
  ]{aipproc}

\layoutstyle{6x9}

\begin{document}

\title{Chromomagnetic Instability and Gluonic Phase at Nonzero Temperature}

\classification{12.38.-t, 11.30.Qc, 26.60.+c}
\keywords{Quark matter, Color superconductivity, 
Chromomagnetic instability, Gluonic phase}

\author{O. Kiriyama}
{address={Institut f\"ur Theoretische Physik, J.W. Goethe-Universit\"at, 
D-60438 Frankfurt/Main, Germany}}

\begin{abstract}
We describe the results of recent studies 
of a chromomagnetic instability and a gluonic phase 
in neutral two-flavor quark matter at nonzero temperature.
\end{abstract}

\maketitle

\section{Introduction}
Sufficiently cold and dense quark matter is a color superconductor. 
Color superconducting phases 
at ``moderate'' density, which is relevant for quark matter 
in the interior of compact stars, have a rich phase structure 
and have recently attracted a great deal of interest \cite{CSC}.

Bulk matter in the interior of compact stars 
should be in $\beta$-equilibrium and be neutral 
with respect to electric and color charges. 
In the two-flavor case, these conditions separate 
the Fermi momenta of up and down quarks and, 
as a consequence, the ordinary BCS state (2SC) is not always 
energetically favored over other unconventional states, e.g., 
the gapless two-flavor 
color superconductivity (g2SC) \cite{Shovkovy2003}. 
However, the 2SC/g2SC phases suffer from a chromomagnetic instability, 
indicated by imaginary Meissner masses of some gluons \cite{Huang2004}. 
The instability related to gluons of color 4--7 
occurs when the ratio of the gap 
over the chemical potential mismatch, $\Delta/\delta\mu$, 
decreases below a value $\sqrt{2}$. 
Resolving the chromomagnetic instability 
and clarifying the nature of the true ground state of dense quark matter 
are central issues in the study 
of color superconductivity \cite{LOFF,Giannakis2004,RedRup,Huang2005,
Hong2005,Gorbar2005,Gorbar2005b,Fukush2006,GHMS2006,Hashimoto2006,
KRS2006,Kiri2006,Iida2006,HJZ,Gatto2007,Hashimoto2007,Kiri2007}. 

So far, several candidates for the ground state 
have been proposed \cite{LOFF,Giannakis2004,RedRup,Huang2005,
Hong2005,Gorbar2005,Gatto2007}. 
In this work, we study the gluonic phase (gluonic cylindrical phase II) 
\cite{Gorbar2005} at nonzero temperature.

\section{Free Energy of Gluonic Phase}
\subsection{Model}
In order to study the gluonic phase, 
we use a gauged Nambu--Jona-Lasinio model with massless up and down quarks:
\begin{eqnarray}
{\cal L}=\bar{\psi}(iD\hspace{-7pt}/+\hat{\mu}\gamma^0)\psi
+G_D\left(\bar{\psi}i\gamma_5\varepsilon\epsilon^bC\bar{\psi}^T\right)
\left(\psi Ci\gamma_5\varepsilon\epsilon^b\psi\right)
-\frac{1}{4}F_{\mu\nu}^{a}F^{a\mu\nu},
\end{eqnarray}
where the quark field $\psi$ carries flavor ($i,j=1,\cdots,N_f$
with $N_f=2$) and color ($\alpha,\beta=1,\cdots,N_c$ with $N_c=3$)
indices, $C$ is the charge conjugation matrix;
$(\varepsilon)^{ik}=\varepsilon^{ik}$ and
$(\epsilon^b)^{\alpha\beta}=\epsilon^{b\alpha\beta}$
are the antisymmetric tensors in flavor and color spaces,
respectively. The diquark coupling strength 
in the scalar ($J^P=0^+$) color-antitriplet channel 
is denoted by $G_D$. 
The covariant derivative and the field
strength tensor are defined as
\begin{eqnarray}
D_{\mu}=\partial_{\mu}-igA_{\mu}^{a}T^{a},~
F_{\mu\nu}^{a}=\partial_{\mu}A_{\nu}^{a}-\partial_{\nu}A_{\mu}^{a}
+gf^{abc}A_{\mu}^{b}A_{\nu}^{c}.
\end{eqnarray}
The elements of the diagonal matrix of quark chemical potentials $\hat{\mu}$ 
in $\beta$-equilibrated neutral two-flavor quark matter are given by
\begin{eqnarray}
&&\mu_{ur}=\mu_{ug}=\bar{\mu}-\delta\mu,~
\mu_{dr}=\mu_{dg}=\bar{\mu}+\delta\mu,\nonumber\\
&&\mu_{ub}=\bar{\mu}-\delta\mu-\mu_8,~
\mu_{db}=\bar{\mu}+\delta\mu-\mu_8,
\end{eqnarray}
with
\begin{eqnarray}
\bar{\mu}=\mu-\frac{\delta\mu}{3}+\frac{\mu_8}{3},
~\delta\mu=\frac{\mu_e}{2}.
\end{eqnarray}

In Nambu-Gor'kov space, the inverse full quark propagator
$S^{-1}(p)$ is written as
\begin{eqnarray}
S^{-1}(p)=\left(
\begin{array}{cc}
(S_0^+)^{-1} & \Phi^- \\
\Phi^+ & (S_0^-)^{-1}
\end{array}
\right),
\end{eqnarray}
with
\begin{eqnarray}
&&(S_0^+)^{-1}=\gamma^{\mu}p_{\mu}+(\bar{\mu}-\delta\mu\tau^3)\gamma^0
+g\gamma^{\mu}A_{\mu}^{a}T^{a},\\
&&(S_0^-)^{-1}=\gamma^{\mu}p_{\mu}-(\bar{\mu}-\delta\mu\tau^3)\gamma^0
-g\gamma^{\mu}A_{\mu}^{a}T^{aT},
\end{eqnarray}
and
\begin{eqnarray}
\Phi^- = -i\varepsilon\epsilon^b\gamma_5\Delta,~
\Phi^+ = -i\varepsilon\epsilon^b\gamma_5\Delta.
\end{eqnarray}
Here $\tau^3=\mbox{diag}(1,-1)$ is a matrix in flavor space. 

For the gluonic phase, $B=\langle gA_z^6 \rangle$ 
is the most relevant vector condensate, 
because the tachyonic mode related to gluons 4--7 exists 
in the direction of $B$ \cite{Gorbar2005,KRS2006}. 
Besides $B$, to ensure color neutrality at $B \neq 0$ 
we have to introduce $\mu_3=\langle gA_0^3 \rangle$ \cite{Gorbar2005}. 
In what follows, however, we neglect 
color chemical potentials $\mu_8$ and $\mu_3$, 
since their effect on the free energy 
is negligibly small for $\alpha_s \simeq 1$. 
Taking account of $B$, the free energy of the gluonic phase 
in the one-loop approximation is given by
\begin{eqnarray}
\Omega(\Delta,\mu_e,B;\mu,T)&=&-\frac{1}{12\pi^2}
\left(\mu_e^4+2\pi^2T^2\mu_e^2+\frac{7\pi^4}{15}T^4\right)\nonumber\\
&&+\frac{\Delta^2}{4G_D}
-\frac{1}{2}\sum_{a}\int\frac{d^3p}{(2\pi)^3}
\left[|\epsilon_a|+2T\ln(1+e^{-\beta|\epsilon_a|})\right],\label{eqn:ep1}
\end{eqnarray}
where $\beta=1/T$, the $\epsilon_a$'s are quasi-quark energies
and the sum runs over all particle and anti-particle $\epsilon_a$'s. 
Note that the $\epsilon_a$'s depend on $B$ through the covariant derivatives 
in the quark propagator. 
Here, we added a contribution from electrons (first line on the r.h.s.). 
In order to evaluate loop diagrams we use a three-momentum cutoff 
$\Lambda=653.3~{\rm MeV}$ throughout this work. 
In addition, in order to remove a cutoff dependence of the free energy 
we introduce the following subtraction:
\begin{eqnarray}
\Omega_R=\Omega(\Delta,\delta\mu,B;\mu,T)-\Omega(0,0,B;0,0).
\end{eqnarray}
Note that this free energy subtraction is not adequate 
to remove the cutoff dependence of the free energy at $T>0$ \cite{Kiri2006}. 
However, we have carefully checked that it is nothing but a cutoff artifact 
and moreover is negligibly small 
at $\mu \sim 400~{\rm MeV}$ 
and at temperatures of interest (20 MeV at most). 
In order to find the gluonic phase, 
we first solve a set of coupled equations,
\begin{eqnarray}
\frac{\partial\Omega_R}{\partial\Delta}
=\frac{\partial\Omega_R}{\partial\delta\mu}=0,
\end{eqnarray}
as a function of $B$ and, then, compute $\Omega_R(B)$. 
Finally, the minimum of $\Omega_R(B)$ determines the gluonic phase.

\subsection{Numerical results}
\begin{figure}
\includegraphics[height=.45\textwidth]{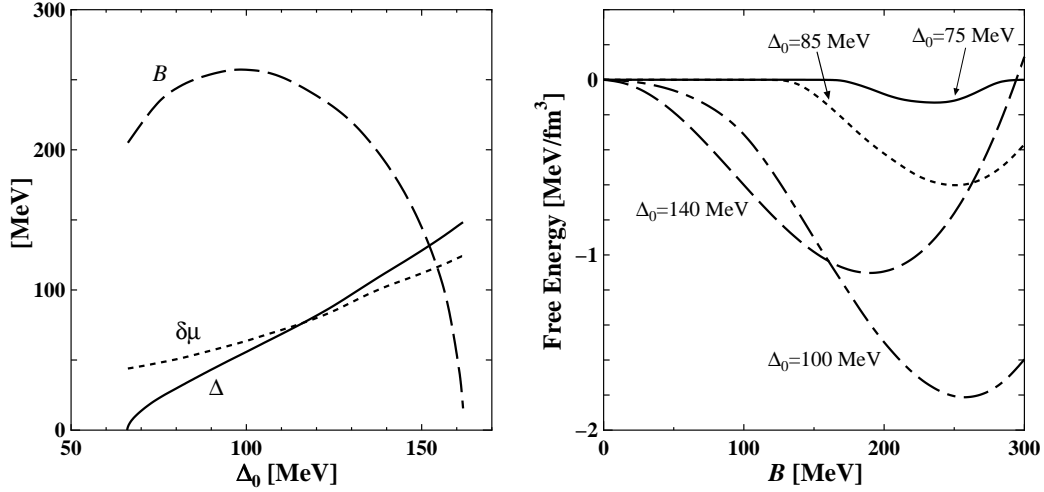}
\caption{Left: The $\Delta_0$ dependence of the gap $\Delta$, 
the chemical potential mismatch $\delta\mu$, 
and the vector condensate $B$ at $T=0$. 
Right: The free energy (measured with respect to the 2SC/g2SC/NQ phases) 
as a function of $B$ at $T=0$. 
The results are plotted for $\mu=400~{\rm MeV}$.}
\label{Fig1}
\end{figure}
In the left panel of Fig. \ref{Fig1}, 
we plot $\Delta$, $\delta\mu$ and $B$ 
as a function of $\Delta_0$, the 2SC gap at $\delta\mu=0$. 
(Essentially, $\Delta_0$ corresponds to the diquark coupling strength.) 
We find that the gluonic phase exists 
in the window $66~{\rm MeV}<\Delta_0<162~{\rm MeV}$ 
and that a strongly (weakly) first-order transition 
between the gluonic phase and the NQ (2SC) phase takes place 
at $\Delta_0=66~(162)~{\rm MeV}$. 
Note that the results are consistent with those in Ref. \cite{Hashimoto2007}, 
where the color chemical potentials $\mu_3$ and $\mu_8$ 
were treated self-consistently. 
The right panel of Fig. \ref{Fig1} shows 
the free energy $\Omega_R(B)$ 
(measured with respect to the 2SC/g2SC/NQ phases at $B=0$) 
as a function of $B$ at $T=0$. 
One can see that, in the cases of $\Delta_0=75,~85~{\rm MeV}$, 
the gluonic phase is energetically more favored 
than the chromomagnetically stable NQ phase at $B=0$. 
[For small values of $B$, we found that the system 
is in the ungapped phase and the free energy 
behaves like $\Omega_R(B) \sim O(B^4)$.] 
On the other hand, for $\Delta_0=100~{\rm MeV}$ 
and $\Delta_0=140~{\rm MeV}$, the 2SC/g2SC phases at $B=0$ are unstable 
and, then, the gluonic phase is energetically favored, as expected. 

\begin{figure}
\includegraphics[height=.45\textwidth]{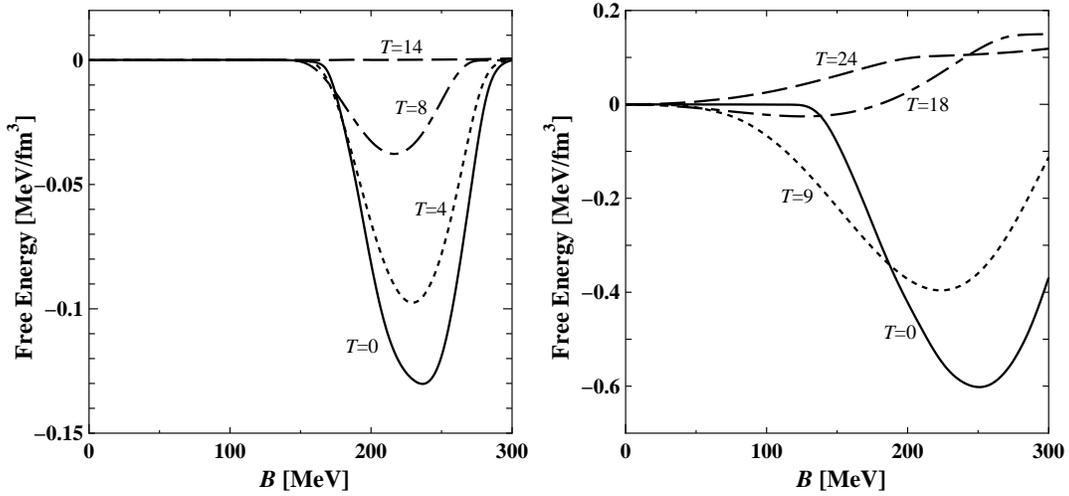}
\caption{The temperature dependence of the free energy as a function of $B$ 
for $\Delta_0=75~{\rm MeV}$ (left) and 
$\Delta_0=85~{\rm MeV}$ (right). 
The quark chemical potential is taken to be $\mu=400~{\rm MeV}$.}
\label{Fig2}
\end{figure}

In Fig. \ref{Fig2}, we plot the temperature dependence of the free energy 
for $\Delta_0=75~{\rm MeV}$ (left) and $\Delta_0=85~{\rm MeV}$ (right). 
At $\Delta_0=75~{\rm MeV}$, the free energy gain gets rapidly reduced 
when $T$ is increased, but the vacuum expectation value of $B$ 
is only slightly reduced. 
Then we find a strongly first-order transition at $T \simeq 14~{\rm MeV}$. 
In the case of $\Delta_0=85~{\rm MeV}$, 
the gluonic phase is energetically more favored 
than the chromomagnetically stable NQ phase at low temperature. 
At $T \simeq 9~{\rm MeV}$, the stable NQ phase undergoes 
a phase transition into the unstable g2SC phase. 
The gluonic phase is energetically favored over 
the unstable g2SC phases until the temperature reaches 
$T \simeq 20~{\rm MeV}$. 
Above this temperature, the gapless dispersion relation 
is smoothed out and the stable g2SC phase is favored. 

Here, we would like to make a comment regarding 
the order of the phase transitions. 
In the weak coupling regime, 
the transition from the gluonic phase to the NQ phase 
is strongly of first order. 
On the other hand, the phase transitions from the gluonic phase 
to the 2SC/g2SC phases in the intermediate coupling regime 
are likely to be of second order. 
However, evaluating the free energy near the critical temperatures 
self-consistently is not easy. 
Thus we cannot exclude the possibility of weakly first-order transitions. 

\section{Conclusion}
\begin{figure}
\includegraphics[height=.45\textwidth]{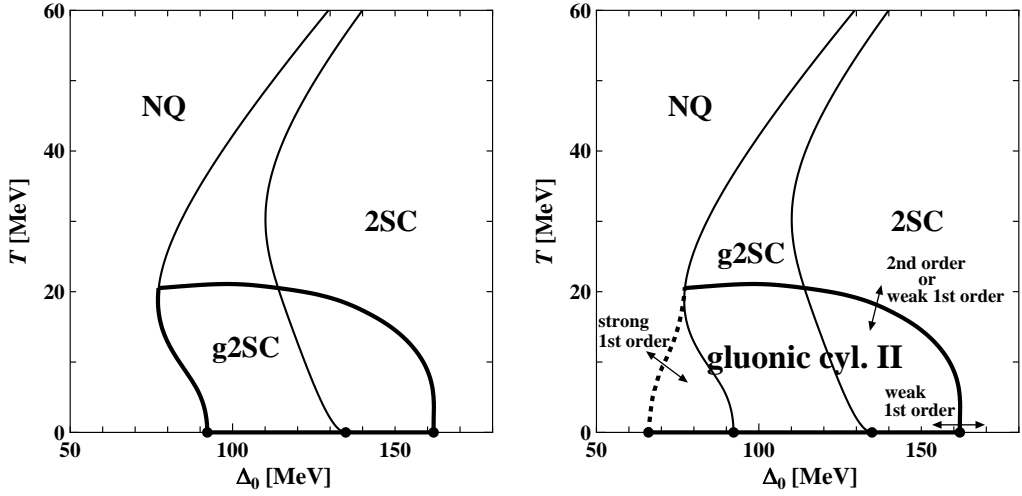}
\caption{Left: The $\Delta_0$-$T$ phase diagram of 
electrically neutral two-flavor quark matter. 
At $T=0$, the g2SC phase phase exists in the window 
$92~{\rm MeV}<\Delta_0<134~{\rm MeV}$ 
and the 2SC window is given by $\Delta_0>134~{\rm MeV}$. 
The unstable region, where gluons 4--7 are tachyonic, 
is depicted by the region enclosed by the thick solid line. 
The g2SC phase and a part of the 2SC phase 
($92~{\rm MeV}<\Delta_0<162~{\rm MeV}$) suffer 
from the chromomagnetic instability at $T=0$. 
Right: Schematic phase diagram of the gluonic phase \cite{Kiri2007}. 
In the region enclosed by the thick solid and dotted lines, 
the gluonic phase is more favored than the 2SC/g2SC/NQ phases. 
First-order phase boundary is indicated by the thick dotted line. 
The thick solid line denotes the line of second-order 
or weakly first-order transitions.}
\label{Fig3}
\end{figure}
In the left panel of Fig. \ref{Fig3}, 
we plot the phase diagram of the electrically neutral 2SC/g2SC phases 
in the plane of $\Delta_0$ versus $T$. 
The region enclosed by the thick solid line is unstable 
because the Meissner mass of gluons 4--7 is imaginary there. 
Note, however, that we did not examine the global structure 
of the free energy self-consistently. 
Therefore we should regard this result 
as the tendency toward the gluonic phase in the 2SC/g2SC phases. 
In fact, at $T=0$, the gluonic cylindrical phase II exists 
in the wider window $66~{\rm MeV}<\Delta_0<162~{\rm MeV}$ 
(see also Ref. \cite{Hashimoto2007}). 

In this work, we studied the phase structure of 
the gluonic cylindrical phase II at nonzero temperature. 
In the weak coupling regime, we found that the phase transition 
between the gluonic phase and the NQ phase is of strongly first order. 
In contrast, in the intermediate and strong coupling regimes, 
the transition is of second order or weakly first order. 
Therefore, in these regimes, the critical line shown 
in the left panel of Fig. \ref{Fig3} is not altered much 
by the self-consistent analysis. 
In the right panel of Fig. \ref{Fig3}, 
we show the schematic phase diagram of the gluonic phase. 
One can see that the gluonic phase 
wins against a part of the NQ phase and enlarges its region. 
The phase transition to the NQ phase is of strongly first order (dotted line) 
while the transition to the 2SC/g2SC phases is of second order 
or weakly first order (solid line). 
The critical temperature for the gluonic phase 
is roughly given by $T_c \sim 20~{\rm MeV}$. 
(It is known that the gluonic phase survives 
until the temperature reaches $T_c \sim \delta\mu^{{\rm(NQ)}}/2$, 
where $\delta\mu^{{\rm (NQ)}}$ denotes the chemical potential mismatch 
in the NQ phase at $T=0$ \cite{Kiri2006}. 
The present model parameters yield $\delta\mu^{{\rm (NQ)}} \sim 40~{\rm MeV}$ 
and, hence, $T_c \sim 20~{\rm MeV}$.) 

Finally, we comment on the outlook for future studies. 
Several candidates for the true ground state 
(for instance, other types of the gluonic phase \cite{Gorbar2005} 
and a LOFF state with multiple plane waves \cite{LOFF}) have been proposed. 
It is then quite interesting to study the free energy of these phases 
and their stability against the chromomagnetic instability. 
In addition, since the gluonic phase dominates the low-temperature region 
of the phase diagram for the wide range of the coupling strength, 
it would have important implications for the properties of compact stars. 

\begin{theacknowledgments}
I would like to thank Dirk Rischke for discussions. 
This work was supported by the Deutsche Forschungsgemeinschaft (DFG).
\end{theacknowledgments}


\begin{thebibliography}{30}
\bibitem{CSC}
K. Rajagopal and F. Wilczek, in
{\it At the Frontier of Particle Physics/Handbook of QCD},
edited by M. Shifman (World Scientific, Singapole, 2001); 
M. G. Alford, Annu. Rev. Nucl. Part. Sci. {\bf 51}, 131 (2001); 
D. K. Hong, Acta. Phys. Pol. B {\bf 32}, 1253 (2001); 
S. Reddy, Acta. Phys. Pol. B {\bf 33}, 4101 (2002); 
D. H. Rischke, Prog. Part. Nucl. Phys. {\bf 52}, 197 (2004); 
R. Casalbuoni and G. Nardulli, Rev. Mod. Phys. {\bf 76}, 263 (2004); 
M. Buballa, Phys. Rept. {\bf 407}, 205 (2005); 
M. Huang, Int. J. Mod. Phys. E {\bf 14}, 675 (2005); 
I. A. Shovkovy, Found. Phys. {\bf 35}, 1309 (2005).

\bibitem{Shovkovy2003}
I. Shovkovy and M. Huang, Phys. Lett. B {\bf 564}, 205 (2003);
M. Huang and I. Shovkovy, Nucl. Phys. {\bf A729}, 835 (2003).

\bibitem{Huang2004}
M. Huang and I.A. Shovkovy, Phys. Rev. D {\bf 70}, 051501(R) (2004);
Phys. Rev. D {\bf 70}, 094030 (2004).

\bibitem{LOFF}M. Alford, J. A. Bowers, and K. Rajagopal, 
Phys. Rev. D {\bf 63}, 074016 (2001); 
J. A. Bowers and K. Rajagopal, Phys. Rev. D {\bf 66}, 065002 (2002).

\bibitem{Giannakis2004}
I. Giannakis and H. C. Ren,
Phys. Lett. B {\bf 611}, 137 (2005);
Nucl. Phys. {\bf B723}, 255 (2005);
I. Giannakis, D. f. Hou, and H. C. Ren,
Phys. Lett. B {\bf 631}, 16 (2005).

\bibitem{RedRup}
S. Reddy and G. Rupak, Phys. Rev. C {\bf 71}, 025201 (2005); 
I. Shovkovy, M. Hanauske, and M. Huang, Phys. Rev. D {\bf 67}, 103004 (2005).

\bibitem{Huang2005}
M. Huang, Phys. Rev. D {\bf 73}, 045007 (2006).

\bibitem{Hong2005}
D. K. Hong, hep-ph/0506097.

\bibitem{Gorbar2005}
E. V. Gorbar, M. Hashimoto, and V. A. Miransky,
Phys. Lett. B {\bf 632}, 305 (2006);
Phys. Rev. D {\bf 75}, 085012 (2007).

\bibitem{Gorbar2005b}
E. V. Gorbar, M. Hashimoto, and V. A. Miransky,
Phys. Rev. Lett. {\bf 96}, 022005 (2006).

\bibitem{Fukush2006}
K. Fukushima, Phys. Rev. D {\bf 73}, 094016 (2006).

\bibitem{GHMS2006}E. V. Gorbar, M. Hashimoto and V. A. Miransky,
and I. A. Shovkovy, Phys. Rev. D {\bf 73}, 111502(R) (2006).

\bibitem{Hashimoto2006}M. Hashimoto, Phys. Lett. B {\bf 642}, 93 (2006).

\bibitem{KRS2006}O. Kiriyama, D. H. Rischke, and I. A. Shovkovy,
Phys. Lett. B {\bf 643}, 331 (2006).

\bibitem{Kiri2006}O. Kiriyama, Phys. Rev. D {\bf 74}, 074019 (2006); 
{\it ibid.} {\bf 74}, 114011 (2006).

\bibitem{Iida2006}
K. Iida and K. Fukushima, Phys. Rev. D {\bf 74}, 074020 (2006).

\bibitem{HJZ}L. He, M. Jin, and P. Zhuang,
Phys. Rev. D {\bf 75}, 036003 (2007).

\bibitem{Gatto2007}R. Gatto and M. Ruggieri,
Phys. Rev. D {\bf 75}, 114004 (2007).

\bibitem{Hashimoto2007}M. Hashimoto and V. A. Miransky, 
Prog. Theor. Phys. {\bf 118}, 303 (2007).

\bibitem{Kiri2007}O. Kiriyama, arXiv:0709.1083 [hep-ph].

\end{thebibliography}
\end{document}